%% file: review.tex
\documentclass[11pt,a4paper]{article}

\usepackage[utf8]{inputenc}
\usepackage{lmodern}
\usepackage{amsmath}
\usepackage{amssymb}
\usepackage{graphicx}
\usepackage{booktabs}
\usepackage{longtable}
\usepackage{cite}
\usepackage{microtype}
\usepackage[colorlinks=true, linkcolor=blue, citecolor=blue, urlcolor=blue]{hyperref}

\usepackage[margin=1.0in]{geometry}

\graphicspath{{./}}

\newcounter{clinbox}

\title{\textbf{Sensory Restoration via Brain-Computer Interfaces: A Scoping Review}}
\author{\textbf{Xuan-The Tran} \\
HAISmartlink Lab\\
\textit{ANCHISTE, Hai Phong, Vietnam}
}
\date{\today}

\begin{document}

\maketitle

\begin{abstract}
\noindent Brain-computer interfaces (BCIs) offer a promising avenue for restoring sensory and motor function in individuals with severe neurological impairment, but the literature is highly fragmented between invasive neuroprosthetics and non-invasive electrophysiological decoders, with inconsistent terminology and comparison metrics. This scoping review maps the landscape of BCI-mediated sensory restoration along a unified $2\times2$ framework defined by invasiveness and signal direction, charts representative modalities and their reported trade-offs, and synthesizes a convergence roadmap for the field. Eligible sources comprised peer-reviewed primary research, clinical trials, and authoritative reviews or perspectives on BCI or neuroprosthetic systems for sensory or motor restoration, substitution, or augmentation, spanning both signal directions, in human or translational preclinical studies published in English between 1969 and 2025; eligibility was deliberately restricted to high-impact venues to prioritize landmark evidence. Rather than an exhaustive database search, we charted a purposively assembled, citation-chained corpus of pivotal sources; thirty-one sources met the criteria and were charted for modality, signal type, invasiveness, signal direction, spatial-temporal resolution, clinical risk, relative cost, and regulatory maturity. We define and distinguish restoration, substitution, and augmentation, and organize the charted sources across the four quadrants of the framework. The corpus is dominated by efferent restoration (21 of 31 sources) and by invasive interfaces (22 of 31), and is concentrated after 2015 (25 of 31). Non-invasive, AI-augmented silent-speech decoding has matured rapidly since 2023, while invasive speech and motor neuroprostheses have achieved near-conversational communication rates. The unified taxonomy clarifies trade-offs between invasive and non-invasive pathways and the role of foundation models in closing the gap between them. We outline a near-, medium-, and long-term roadmap toward closed-loop, bidirectional restoration, and identify gaps in metric standardization, longitudinal evidence, and cross-community collaboration as priorities for future research.
\end{abstract}

\vspace{2mm}

\newpage

\section{Introduction}
\label{sec:intro}

\subsection{Rationale}
\label{subsec:rationale}
For individuals with late-stage neurodegenerative disorders such as amyotrophic lateral sclerosis (ALS), as well as those affected by stroke, spinal cord injury, or severe trauma, the loss of motor control constitutes a multi-faceted disruption of both physical autonomy and sensory-mediated communication with the environment. Globally, the prevalence of motor and sensory deficits arising from traumatic brain injury, spinal cord lesions, cortical strokes, and neurodegenerative conditions affects millions of individuals annually, presenting a profound societal and clinical challenge. While brain-computer interfaces (BCIs) offer a revolutionary pathway toward restoring these lost faculties \cite{lebedev2017brain, chaudhary2016brain}, the scientific literature remains highly fragmented. Research is largely bifurcated into invasive neuroprosthetics focused on direct cortical stimulation and recording, and non-invasive electroencephalography (EEG) or functional near-infrared spectroscopy (fNIRS) paradigms designed for motor and speech decoding. This compartmentalization, compounded by inconsistent terminology---where ``restoration,'' ``substitution,'' and ``augmentation'' are frequently used interchangeably---obscures the fundamental physiological and engineering trade-offs inherent to each approach.

\subsection{Existing Reviews and the Conceptual Gap}
\label{subsec:gap}
Existing surveys treat invasive neuroprosthetics and non-invasive EEG-BCI as separate literatures, leaving the reader without a unified framework to compare trade-offs across both pathways. Prior reviews typically silo these domains; for instance, surveys focused on invasive cortical interfaces (e.g., \cite{fernandez2021visual, hochberg2012reach}) rarely address non-invasive speech decoding advancements, while EEG-centric reviews (e.g., \cite{wolpaw2002brain, chaudhary2016brain}) disregard direct microstimulation paradigms. This polarization creates two primary limitations. First, it introduces terminological confusion, where terms such as \textit{restoration}, \textit{substitution}, and \textit{augmentation} are used interchangeably, obscuring whether a system restores a native sensory pathway or merely bypasses it via an alternative modality. Second, existing literature is largely obsolete regarding the rapid convergence of deep learning and brain-computer interfaces. Most reviews predate the 2023--2025 breakthroughs in foundation models and self-supervised alignment that made non-invasive silent-speech and imagined-speech decoding clinically practical. Consequently, researchers lack a cohesive roadmap to navigate the clinical and engineering trade-offs of these rapidly evolving paradigms. These gaps underscore the necessity of a unified structural treatment of the field.

\subsection{A Unified $2\times2$ Conceptual Framework}
\label{subsec:framework_intro}
The essence of sensory restoration is fundamentally structured by a two-dimensional question: through which physical interface, and in which direction of information flow, does the neural communication occur? The signal direction dictates the computational and biophysical constraints of the BCI system. Efferent pathways (sensory-OUT)---such as motor control and speech neuroprosthetics---rely on reading and decoding neural intent, which presents a lower risk profile and has already achieved clinical viability. In contrast, afferent pathways (sensory-IN)---including visual, auditory, and somatosensory cortical prostheses---require writing information directly into neural tissue via electrical microstimulation or neuromodulation. This afferent channel carries a significantly higher regulatory and safety barrier. While non-invasive interfaces are physically constrained from delivering high-density cortical writing, they can achieve coarse sensory-IN through sensory substitution (e.g., tactile-to-visual mapping) or non-invasive neuromodulation techniques such as transcranial focused ultrasound (tFUS) and transcranial alternating current stimulation (tACS). The primary goal of this review is to apply a unified $2\times2$ framework---defined by invasiveness and signal direction---that establishes precise taxonomic boundaries (restoration, substitution, and augmentation) and maps out a concrete convergence roadmap for both pathways.

\subsection{Review Objectives}
\label{subsec:objectives}
Constructing a unified framework that spans these diverse domains is complicated by three core challenges. First, the communities developing invasive and non-invasive BCIs employ divergent metrics, terminology, and evaluation norms. Comparing the bits-per-minute throughput of a motor-imagery EEG speller with the phosphene count of a visual cortical implant is non-trivial, rendering simple flat reviews ineffective. Second, the boundary between sensory input and output is not strictly binary; hybrid and closed-loop systems, as well as non-invasive sensory substitution methods, sit in a graded spectrum that resists rigid categorization. Third, the rapid acceleration of state-of-the-art benchmarks (e.g., the UCSF BRAVO 2023 speech avatar \cite{metzger2023high}, BrainGate 2023 high-performance speech decoding \cite{willett2023high}, and Meta's 2023 MEG speech perception decoding \cite{defossez2023decoding}) means that any review relying solely on numeric milestones is rendered obsolete immediately upon publication. To remain durable, this review focuses on the fundamental structural principles and physical constraints that govern these technologies.

\begin{figure}[htbp]
\centering
\includegraphics[width=0.9\textwidth]{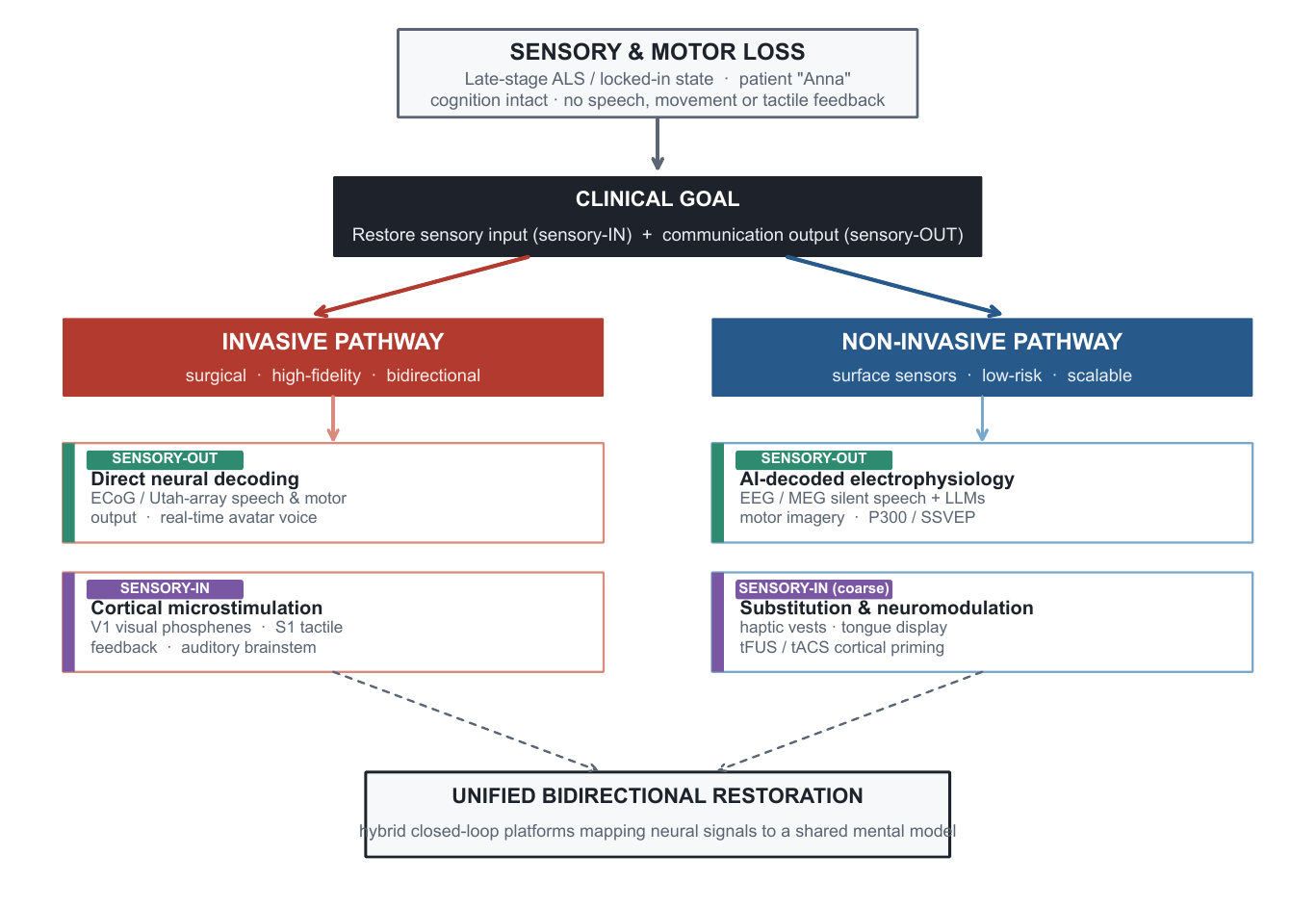}
\caption{Architectural bifurcation of sensory restoration. Starting from clinical sensory and motor pathology (such as in late-stage ALS or a locked-in state; Box~\ref{box:anna}), clinical interventions split into the invasive pathway (direct surgical implantation of Utah arrays, ECoG grids, or endovascular stents providing high-fidelity, bidirectional pathways) and the non-invasive pathway (scalp EEG/MEG recording and sensory substitution offering scalable, lower-risk communication channels).}
\label{fig:pathways}
\end{figure}

We address these challenges through three organizing devices designed to synthesize the fragmented literature into a cohesive, navigable map. First, we apply a unified taxonomy and a comprehensive trade-off table (\S\ref{sec:comparison}) that evaluates systems across shared dimensions of spatial-temporal resolution, invasiveness, regulatory maturity, and financial cost, establishing an apples-to-apples comparison framework. Second, we present a graded $2\times2$ matrix that maps representative systems along a continuum, explicitly positioning coarse sensory-IN approaches like haptic substitution and neuromodulation in a transitional zone between non-invasive and invasive afferent pathways. Third, rather than focusing on ephemeral benchmark scores, we present a principle-first convergence roadmap organized by durable physiological and physical limits, mapping two distinct trajectories: a non-invasive, AI-driven silent-speech BCI that can restore communication utility in the immediate term, and a bidirectional invasive cortical implant that offers a long-term pathway toward genuine sensory restoration, with these parallel tracks eventually merging. The structural bifurcation of this clinical and technical landscape is illustrated in Figure~\ref{fig:pathways}.

In summary, this review pursues four objectives:
\begin{enumerate}
    \item \textbf{Apply a unified $2\times2$ taxonomy} (invasiveness $\times$ signal direction) with precise definitions distinguishing sensory restoration, substitution, and augmentation, providing a common vernacular across clinical and engineering fields (\S\ref{subsec:taxonomy}).
    \item \textbf{Chart the four quadrants} of the BCI landscape, culminating in a cross-pathway trade-off table evaluating safety, resolution, cost, and maturity (\S\ref{subsec:invasive}--\S\ref{sec:comparison}).
    \item \textbf{Synthesize a multi-phase convergence roadmap} detailing the transition from near-term non-invasive speech decoding to medium-term closed-loop implants, and ultimately long-term bidirectional, AI-fused cortical interfaces (\S\ref{subsec:roadmap}).
    \item \textbf{Identify ethical and regulatory considerations} relevant to sensory neuroprosthetics, addressing neural privacy, user identity, cognitive autonomy, and regulatory pathways (\S\ref{subsec:ethics}).
\end{enumerate}

Through this structured synthesis, we show that for patients with severe sensory and motor impairment (Box~\ref{box:anna}), the core question is no longer whether sensory restoration is possible, but rather which technical pathway is optimal under specific clinical and personal constraints.

\begin{center}
\refstepcounter{clinbox}\label{box:anna}
\fbox{\begin{minipage}{0.92\textwidth}
\vspace{1mm}
\textbf{Box \theclinbox. Illustrative clinical vignette --- ``Anna.''}

\noindent Throughout the Discussion and Conclusion, we use a hypothetical patient, ``Anna,'' a woman in a locked-in state (LIS) following late-stage ALS, as a single recurring illustrative example for applying the $2\times2$ framework. Although Anna's cognitive functions remain entirely preserved, she is functionally isolated: unable to articulate speech, generate voluntary motor responses, or receive tactile validation of external interactions. Anna's case is referenced where the framework's abstract trade-offs (Table~\ref{tab:trade_offs}, Figures~\ref{fig:framework_2x2}--\ref{fig:convergence_roadmap}) translate into a concrete clinical trajectory: from an immediate, non-invasive silent-speech BCI, through a medium-term implanted speech neuroprosthesis, to a long-term bidirectional system restoring both communication and sensation (\S\ref{subsec:roadmap}, \S\ref{sec:conclusion}).
\vspace{1mm}
\end{minipage}}
\end{center}

\section{Methods}
\label{sec:methods}
\textit{This section describes the scoping-review methodology (PRISMA-ScR) used to identify, screen, and chart the sources of evidence underlying the taxonomy and synthesis presented in \S\ref{sec:results}--\S\ref{sec:discussion}.}

\subsection{Protocol and Design}
\label{subsec:protocol}
This work is a structured scoping review conducted as a purposive synthesis of landmark evidence. It follows the PRISMA Extension for Scoping Reviews (PRISMA-ScR) reporting checklist \cite{tricco2018prisma} for the reporting items applicable to a curated evidence base. No protocol was registered in advance; this is acknowledged as a limitation (\S\ref{sec:limitations}). Consistent with scoping-review methodology, the objective is to map the conceptual landscape and chart the characteristics of representative sources, not to perform a quantitative meta-analysis or a formal risk-of-bias appraisal of individual studies. We note at the outset that, by design, the search was not exhaustive: eligibility was deliberately restricted to high-impact peer-reviewed venues in order to anchor the synthesis in pivotal, well-validated milestones (\S\ref{subsec:search}, \S\ref{sec:limitations}).

\subsection{Eligibility Criteria}
\label{subsec:eligibility}
Eligibility was defined using the Population--Concept--Context (PCC) framework recommended for scoping reviews (Table~\ref{tab:pcc}).

\begin{table}[htbp]
\centering
\caption{Eligibility criteria (Population--Concept--Context)}
\label{tab:pcc}
\begin{tabular}{p{0.18\textwidth}p{0.74\textwidth}}
\toprule
\textbf{Element} & \textbf{Criteria} \\
\midrule
Population & Humans (and translational animal/preclinical studies directly informing human BCI translation) with sensory and/or motor impairment due to neurological injury, disease, or congenital condition. \\
Concept & Brain-computer interface or neuroprosthetic systems intended for sensory/motor \textit{restoration}, \textit{substitution}, or \textit{augmentation}, spanning afferent (sensory-IN) and efferent (sensory-OUT) signal directions. \\
Context & Peer-reviewed primary research, clinical trials, and authoritative review/perspective articles covering both invasive (intracortical, ECoG, endovascular) and non-invasive (EEG, MEG, fNIRS, sEMG, non-invasive neuromodulation, sensory substitution) interface types. English-language publications from 1969 to 2025, restricted to high-impact venues (predominantly the \textit{Nature}, \textit{Science}, and \textit{Cell} families and leading clinical journals such as \textit{NEJM}, \textit{Lancet}, and \textit{JAMA Neurology}; see \S\ref{subsec:search}). \\
\bottomrule
\end{tabular}
\end{table}

\subsection{Information Sources and Identification Strategy}
\label{subsec:search}
Sources were identified through a purposive, citation-driven strategy rather than an exhaustive multi-database search. Starting from a set of seminal milestones spanning the four quadrants of the framework---for example, the first intracortical control in tetraplegia \cite{hochberg2006neuronal}, visual cortical stimulation in a blind volunteer \cite{fernandez2021visual}, real-time speech neuroprostheses \cite{moses2021neuroprosthesis, metzger2023high}, and non-invasive speech decoding \cite{defossez2023decoding}---we performed backward and forward citation chaining to locate the pivotal primary studies, clinical trials, and authoritative reviews that define each quadrant. Candidate sources were then filtered against the eligibility criteria of Table~\ref{tab:pcc}, with the deliberate restriction to high-impact peer-reviewed venues described in \S\ref{subsec:protocol}. This identification strategy favors breadth of conceptual coverage and well-validated, landmark evidence over exhaustive enumeration; the trade-off is an acknowledged selection bias toward high-profile results, examined in \S\ref{sec:limitations}. The conceptual search space corresponds to the Boolean expression (``brain-computer interface'' OR ``brain-machine interface'' OR neuroprosthe* OR ECoG OR intracortical OR EEG OR MEG) AND (``sensory restoration'' OR ``sensory substitution'' OR ``speech decoding'' OR ``motor imagery'' OR phosphene OR microstimulation OR neuromodulation), which may be used to extend this corpus in future updates.

\subsection{Selection of Sources of Evidence}
\label{subsec:selection}
\begin{figure}[htbp]
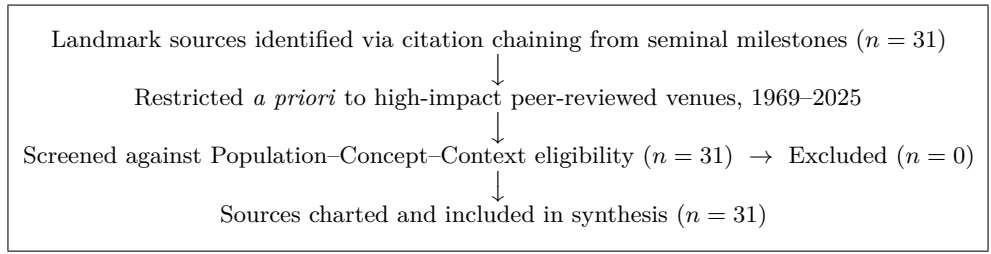

\centering
\fbox{\begin{minipage}{0.8\textwidth}
\centering
\vspace{2mm}
\footnotesize
Landmark sources identified via citation chaining from seminal milestones ($n = 31$) \\
$\big\downarrow$ \\
Restricted \textit{a priori} to high-impact peer-reviewed venues, 1969--2025 \\
$\big\downarrow$ \\
Screened against Population--Concept--Context eligibility ($n = 31$) $\;\rightarrow\;$ Excluded ($n = 0$) \\
$\big\downarrow$ \\
Sources charted and included in synthesis ($n = 31$)
\vspace{2mm}
\end{minipage}}
\caption{Identification and charting flow for the purposive scoping review. Following PRISMA-ScR principles but adapted to a curated evidence base: sources were identified by citation chaining from seminal milestones, restricted \textit{a priori} to high-impact venues, and screened against the eligibility criteria of Table~\ref{tab:pcc}. The single eligibility-confirmation stage (rather than a large screening funnel) reflects the deliberately restricted scope (\S\ref{subsec:protocol}, \S\ref{sec:limitations}).}
\label{fig:prisma}
\end{figure}

Candidate sources identified by citation chaining (\S\ref{subsec:search}) were screened by a single reviewer against the Population--Concept--Context criteria of Table~\ref{tab:pcc}. Because the corpus was purposively pre-filtered to landmark studies in high-impact venues, screening functioned as eligibility confirmation rather than large-scale exclusion: all 31 candidate sources met the criteria and were retained for charting. The single-reviewer process and the absence of a large excluded pool are limitations of this design (\S\ref{sec:limitations}). The identification-to-charting flow is summarized in Figure~\ref{fig:prisma}.

\subsection{Data Charting Process}
\label{subsec:charting}
For each included source, we charted: (i) the BCI modality or platform (e.g., Utah array, ECoG grid, scalp EEG, tFUS); (ii) signal type (e.g., single-unit spikes, local field potentials, scalp voltages, acoustic waves); (iii) invasiveness category; (iv) signal direction (afferent/efferent); (v) spatial and temporal resolution; (vi) clinical risk profile; (vii) relative cost; and (viii) regulatory/maturity stage. These items populate the source-characteristics table (Table~\ref{tab:charted}), the cross-pathway trade-off matrix (Table~\ref{tab:trade_offs}), the $2\times2$ taxonomy map (Figure~\ref{fig:framework_2x2}), and the trade-off bubble chart (Figure~\ref{fig:comparison_plot}). Charting was performed by a single reviewer using a standardized extraction template; no second-reviewer verification was undertaken, which is noted as a limitation (\S\ref{sec:limitations}).

\section{Results}
\label{sec:results}

\subsection{Overview of Charted Sources}
\label{subsec:overview}
The 31 charted sources are summarized individually in Table~\ref{tab:charted} and aggregated in Table~\ref{tab:distribution}. The corpus comprises 26 primary studies (clinical or directly translational), 4 reviews, and 1 ethics perspective; 24 of the primary studies report human in-vivo data, with 2 preclinical platform/feasibility studies.

Three patterns are evident from this distribution. First, the high-impact evidence base is heavily weighted toward \textit{efferent} restoration (21/31 sources) and toward \textit{invasive} interfaces (22/31), reflecting both the greater clinical maturity of read-out neuroprostheses and a publication bias of leading venues toward high-bandwidth invasive demonstrations; afferent sensory write-in (6/31) and non-invasive modalities (5/31) are comparatively under-represented. Second, the evidence is recent and accelerating: 25 of 31 sources were published in 2015 or later, with pronounced clustering in 2016, 2021, and 2023 (five sources each), consistent with the field's rapid, milestone-driven progress (Figures~\ref{fig:invasive_speech_timeline}--\ref{fig:noninvasive_speech_timeline}). Third, by venue, the corpus is concentrated in the \textit{Nature} family (17 sources), the \textit{Science} family (3), \textit{Cell} (1), and leading clinical journals (\textit{NEJM}, \textit{Lancet}, \textit{JAMA Neurology}; 5), with the remainder in specialist venues---a direct consequence of the eligibility restriction (\S\ref{subsec:search}) and a determinant of the selection bias discussed in \S\ref{sec:limitations}.

\begin{table}[htbp]
\centering
\caption{Distribution of the 31 charted sources across charting dimensions.}
\label{tab:distribution}
\vspace{1mm}
\begin{tabular}{llc}
\toprule
\textbf{Dimension} & \textbf{Category} & \textbf{Sources ($n$)} \\
\midrule
Signal direction & Efferent (sensory-OUT) & 21 \\
                 & Afferent (sensory-IN) & 6 \\
                 & Cross-cutting (review/ethics) & 4 \\
\midrule
Invasiveness     & Invasive & 22 \\
                 & Non-invasive & 5 \\
                 & Cross-cutting & 4 \\
\midrule
$2\times2$ quadrant & Invasive afferent & 4 \\
                    & Invasive efferent & 18 \\
                    & Non-invasive efferent & 3 \\
                    & Non-invasive afferent & 2 \\
                    & Cross-cutting (--) & 4 \\
\midrule
Source type      & Primary (clinical/translational) & 26 \\
                 & Review & 4 \\
                 & Perspective & 1 \\
\midrule
Publication period & 1969--2014 & 6 \\
                   & 2015--2025 & 25 \\
\bottomrule
\end{tabular}
\end{table}

\input{charted_sources_table}

\subsection{A Unified Conceptual Framework: The $2\times2$ Taxonomy}
\label{subsec:taxonomy}

\subsubsection{Afferent (Sensory-IN) vs. Efferent (Sensory-OUT)}
The functional architecture of a brain-computer interface (BCI) is fundamentally determined by the direction of information flow relative to the central nervous system. This directional dichotomy splits interfaces into efferent (sensory-OUT) and afferent (sensory-IN) pathways, which differ not only in their clinical objectives but also in their biological constraints and signal-processing paradigms.

Efferent pathways focus on the acquisition and decoding of neural intentions. These systems read endogenous electrical signals generated by motor or cognitive areas of the brain---such as the primary motor cortex (M1) or the ventral motor cortex (vMC)---and translate them into command signals for external effectors. The engineering challenge in efferent BCI is primarily analytical: processing noisy, non-stationary neural signals, extracting descriptive features (such as local field potentials or single-unit spike trains), and deploying machine learning decoders to reconstruct motor or linguistic trajectories. Because efferent systems only read from the brain, they present a lower physiological risk profile, as no external energy is introduced into the neural tissue.

Conversely, afferent pathways (sensory-IN) are designed to write information back into the nervous system. These systems bypass damaged peripheral receptor organs (e.g., the retina or cochlea) and directly stimulate neural structures using electrical current or acoustic waves to evoke artificial sensory percepts. The primary challenge in afferent BCI is biophysical and electrochemical. It requires injecting precise charges into neural tissue without triggering excitotoxicity, glial responses, or electrode degradation. Afferent systems must match the native spatial and temporal coding of sensory pathways to generate coherent perceptions (such as phosphenes or tactile textures). This subjects sensory-IN systems to a significantly higher regulatory and safety barrier compared to their sensory-OUT counterparts.

Comprehensive sensory restoration for a single patient typically requires both pathways in combination: an efferent decoder to read speech and motor intentions, allowing interaction with the environment, and an afferent stimulator to write in somatosensory feedback, closing the loop (Box~\ref{box:anna}).

\subsubsection{The Invasive--Non-Invasive Spectrum}
The physical interface between the device and the nervous system spans a broad spectrum of invasiveness. This spectrum represents a fundamental trade-off between signal-to-noise ratio (SNR) and clinical risk.

At the highly invasive end of the spectrum are intracortical microelectrode arrays (e.g., the Utah Electrode Array), which are surgically implanted directly into the gray matter of the brain. By placing recording and stimulating contacts within micrometers of individual neurons, intracortical arrays provide single-unit resolution (action potentials), enabling high-dimensional control and localized stimulation. However, this proximity comes at the cost of craniotomy, risk of infection, and a chronic tissue response characterized by glial scarring. Moving outward, electrocorticography (ECoG) grids are placed on the surface of the cortex, either sub-durally or epidurally. ECoG bypasses the high-frequency filtering of the skull bone, providing excellent spatial resolution (on the order of millimeters) and high temporal fidelity (capturing local field potentials and gamma-band activity) without penetrating the parenchymal tissue.

Further along the spectrum are minimally invasive interfaces, such as endovascular stent-electrodes, which are navigated through the venous vasculature (e.g., the superior sagittal sinus) to record from adjacent motor areas without direct brain tissue penetration. Finally, non-invasive interfaces reside entirely outside the skull. Scalp electroencephalography (EEG), magnetoencephalography (MEG), and functional near-infrared spectroscopy (fNIRS) record bulk electrical, magnetic, or hemodynamic signals. While non-invasive modalities offer maximum safety, rapid setup, and high accessibility, the skull bone acts as a spatial low-pass filter, severely attenuating high-frequency neural oscillations and scattering spatial details. Consequently, non-invasive systems struggle to achieve high information throughput or localized sensory write-in without sophisticated machine learning reconstruction models.

Clinical pathways often must navigate this spectrum progressively, starting with non-invasive systems for immediate communication needs before transitioning to higher-fidelity invasive interfaces as requirements evolve (Box~\ref{box:anna}).

\subsubsection{The $2\times2$ Map: Intersecting Invasiveness and Information Flow}
By intersecting the axis of signal direction (afferent vs. efferent) with the axis of physical interface depth (invasive vs. non-invasive), we form a unified $2\times2$ taxonomy that maps the heterogeneous BCI landscape into four distinct quadrants. The quadrants are defined as:
\begin{enumerate}
    \item \textbf{Invasive Afferent (Sensory-IN):} Direct cortical stimulation (e.g., visual and somatosensory implants).
    \item \textbf{Invasive Efferent (Sensory-OUT):} Direct cortical decoding (e.g., Utah array motor control, ECoG speech decoders).
    \item \textbf{Non-Invasive Efferent (Sensory-OUT):} Scalp-recorded decoding (e.g., EEG motor imagery, P300 spellers, MEG speech reconstruction).
    \item \textbf{Non-Invasive Afferent (Sensory-IN):} Coarse sensory write-ins (e.g., sensory substitution systems, transcranial ultrasound stimulation).
\end{enumerate}

\begin{figure}[htbp]
\centering
\includegraphics[width=0.7\textwidth]{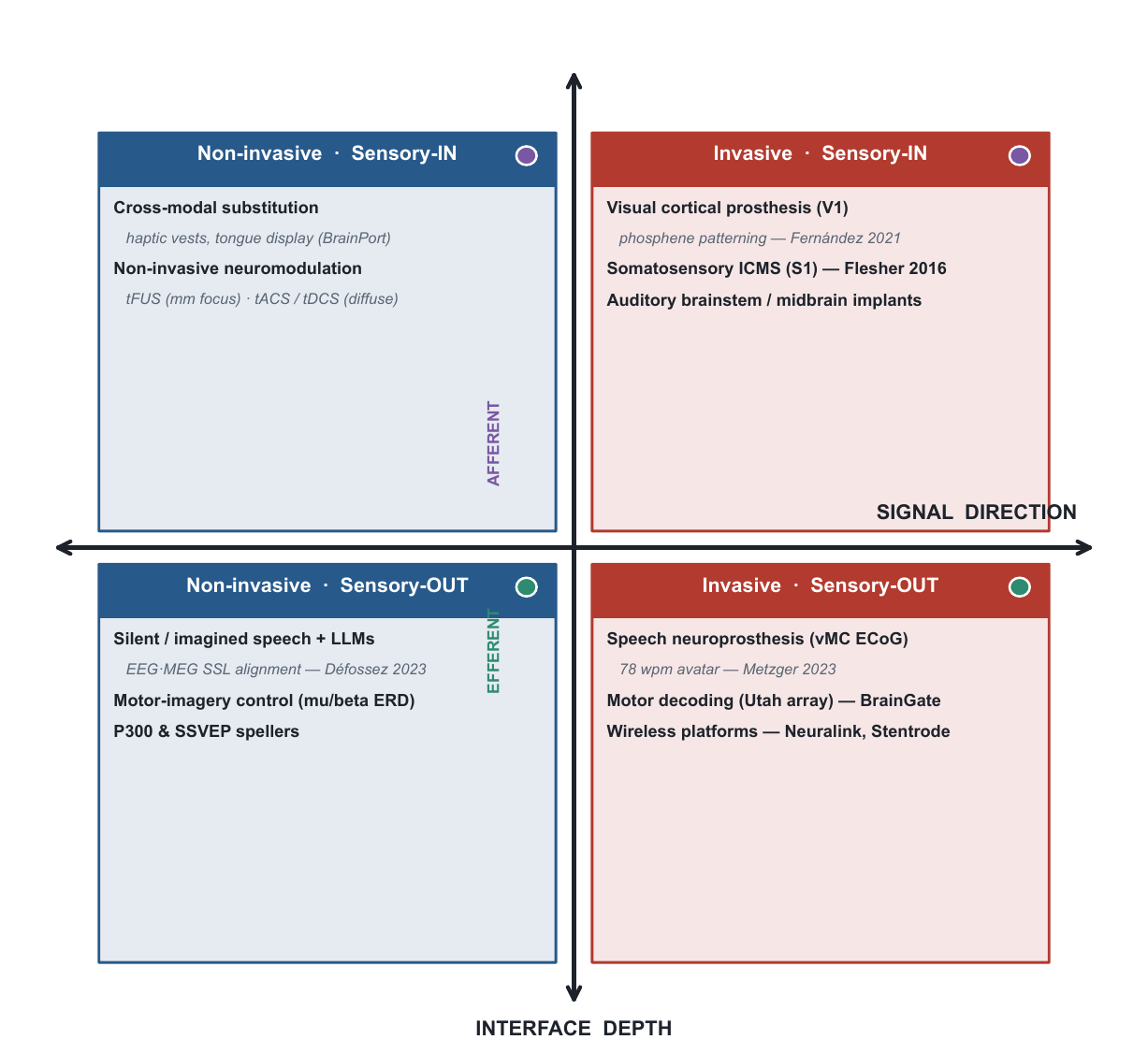}
\caption{The $2\times2$ brain-computer interface (BCI) framework. Modalities are organized along two perpendicular axes: interface depth (invasive vs. non-invasive) and information flow direction (afferent sensory-IN vs. efferent sensory-OUT), creating four quadrants containing representative sensory/motor interfaces.}
\label{fig:framework_2x2}
\end{figure}

Understanding where a clinical application falls within this $2\times2$ map allows researchers to assess the physical limits of the system, the expected information bandwidth, and the translational barriers before starting clinical deployment. This taxonomic organization is visually mapped in Figure~\ref{fig:framework_2x2}.

This map can also be used to categorize an individual patient's therapeutic options and to plan a stepwise transition across quadrants as clinical needs evolve --- e.g., starting from a Quadrant 3 setup (scalp EEG covert-speech speller) in the near term, with a planned transition to a Quadrant 2 setup (implanted ECoG speech neuroprosthesis) as communication needs expand (Box~\ref{box:anna}).

\subsubsection{Defining Restoration, Substitution, and Augmentation}
To prevent conceptual conflation, we define:
\begin{itemize}
    \item \textbf{Restoration:} Re-establishing the original, damaged biological communication pathway (e.g., direct electrical stimulation of V1 to restore functional phosphene sight).
    \item \textbf{Substitution:} Routing sensory information through an alternative sensory channel (e.g., converting camera images into tactile vibration arrays on the skin).
    \item \textbf{Augmentation:} Extending human capabilities beyond baseline physiological ranges (e.g., decoding cognitive speech intent directly or expanding vision to the infrared spectrum).
\end{itemize}

These distinctions matter clinically: restoring native vocal-tract control (restoration) represents a different goal than translating environmental sound into haptic feedback on the skin (substitution), or using predictive language models to auto-complete decoded text (augmentation).

\subsection{Invasive Modalities}
\label{subsec:invasive}
\textit{This subsection charts surgical neural interfaces that record from or microstimulate cortical and subcortical structures directly, providing high-fidelity sensory restoration.}

\subsubsection{Sensory-IN: Visual Cortical Prostheses}
Visual cortical prostheses are designed to restore functional vision to profoundly blind individuals by bypassing damaged ocular structures and directly stimulating the primary visual cortex (V1). The principal mechanism involves evoking phosphenes---perceived points of light in the visual field---via electrical microstimulation of V1 neurons. In a landmark clinical study, Fern\'{a}ndez et al. (2021) \cite{fernandez2021visual} implanted a 100-electrode Utah array into the occipital cortex of a blind human volunteer. By applying microstimulation pulse trains (typically with current amplitudes ranging between 10 to 100~$\mu$A and frequencies between 100 to 200~Hz), the patient successfully perceived and identified simple shapes, letters, and spatial patterns. Complementary work by Beauchamp et al. (2020) \cite{beauchamp2020dynamic} demonstrated that tracing forms through dynamic, sequential stimulation of multiple electrodes enabled both sighted and blind participants to recognize letter shapes, indicating that the temporal choreography of stimulation---not merely the static spatial layout of electrodes---is critical to evoking coherent percepts.

However, translating these achievements into high-resolution vision presents severe engineering hurdles. Evoking a stable, dense visual field requires hundreds of microstimulation sites. As the density of electrodes increases, issues of phosphene fusion, spatial overlap, and cortical phosphene drift arise. Furthermore, the relationship between stimulation current and perceived phosphene brightness is highly non-linear, demanding real-time computational mapping models.

\subsubsection{Sensory-IN: Auditory Brainstem and Midbrain Implants}
When the cochlea or the auditory nerve is damaged---as in cases of neurofibromatosis type II or severe skull base trauma---standard cochlear implants are clinically ineffective. In these cases, auditory brainstem implants (ABIs) and auditory midbrain implants (AMIs) are deployed. ABIs target the cochlear nucleus, while AMIs place multi-contact electrodes in the inferior colliculus. These interfaces utilize the tonotopic organization of these structures, where specific channels are stimulated to represent distinct audio frequencies.

The primary technical challenge is the lower temporal resolution achieved at these brainstem sites compared to the auditory nerve. Consequently, while ABIs and AMIs reliably restore sound envelope detection and assist with lip-reading, the reconstruction of complex frequency components (such as speech pitch and vocal melodies) remains limited.

\subsubsection{Sensory-IN: Somatosensory Cortical Stimulation}
Restoring somatic sensation, particularly the sense of touch and proprioception, is essential for closed-loop control of prosthetic limbs and for emotional communication. Somatosensory implants target the primary somatosensory cortex (S1) in the postcentral gyrus. Flesher et al. (2016) \cite{flesher2016intracortical} demonstrated that intracortical microstimulation (ICMS) of S1 in a human participant could evoke localized tactile sensations---such as pressure, tapping, and vibration---that felt as if they originated from the participant's own paralyzed hand.

By integrating S1 microstimulation with pressure sensors on a robotic hand, Flesher et al. (2021) \cite{flesher2021brain} created a bidirectional closed-loop system in which the restored sense of touch approximately halved the time required to complete object-transfer tasks, as the user no longer had to rely solely on vision to regulate grasp force.

\subsubsection{Sensory-OUT: Motor Neuroprosthetics}
Motor neuroprosthetics focus on decoding movement intentions directly from M1 to restore motor agency. The first demonstration that a person with tetraplegia could operate external devices through an intracortical implant came from Hochberg et al. (2006) \cite{hochberg2006neuronal}; the BrainGate consortium subsequently demonstrated reach and grasp with a robotic arm \cite{hochberg2012reach}, and Collinger et al. (2013) \cite{collinger2013high} achieved high-performance ten-degree-of-freedom control of an anthropomorphic limb by an individual with tetraplegia. The decoding pipeline typically translates firing rates of motor neurons into velocity vectors using Kalman filtering---optimized algorithm design was central to the clinical translation of these systems \cite{gilja2015clinical}---and has recently been augmented by recurrent neural networks (RNNs) for multi-degree-of-freedom control. Motor intent need not be read from primary motor cortex alone: Aflalo et al. (2015) \cite{aflalo2015decoding} decoded reach goals from the posterior parietal cortex, and Bouton et al. (2016) \cite{bouton2016restoring} routed decoded cortical signals directly to functional electrical stimulation of the participant's own forearm muscles, restoring volitional movement of a paralyzed hand.

Willett et al. (2021) \cite{willett2021high} achieved a major milestone in high-performance communication by decoding the neural trajectories of hand movements during imagined handwriting. By aligning these trajectory decoders with a language model, the participant achieved typing speeds of 90 characters per minute with over 99\% accuracy.

\subsubsection{Sensory-OUT: Speech Neuroprosthetics}
For patients who have lost the ability to articulate speech due to paralysis or ALS, speech neuroprosthetics represent the state of the art in efferent communication restoration. Unlike spelling systems that require letter-by-letter selection, speech neuroprostheses decode vocal motor intent directly from the ventral motor cortex (vMC) and Broca's area. This paradigm matured through two pivotal precursors: Anumanchipalli et al. (2019) \cite{anumanchipalli2019speech} synthesized intelligible audible speech by decoding the kinematics of the vocal tract from cortical activity, and Moses et al. (2021) \cite{moses2021neuroprosthesis} achieved the first real-time decoding of full words and sentences from the sensorimotor cortex of a person with anarthria. Building on these foundations, Metzger et al. (2023) \cite{metzger2023high} at UCSF demonstrated an ECoG-based speech neuroprosthesis that translated motor patterns of the vocal tract (jaw, tongue, larynx) into text at 78 words per minute, mapping the decoded intent onto a personalized digital avatar that spoke and emoted in real time.

\begin{figure}[htbp]
\centering
\includegraphics[width=0.7\textwidth]{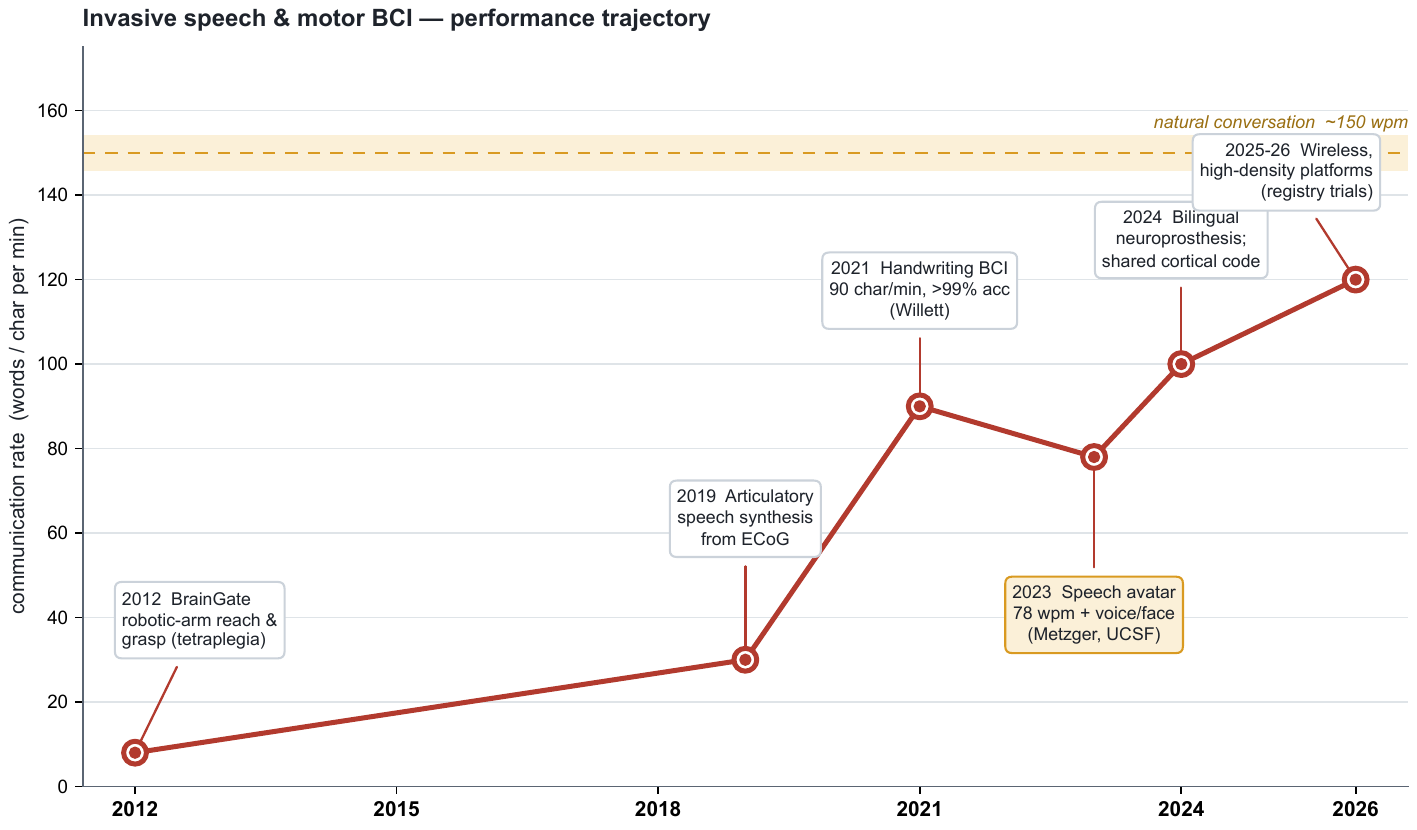}
\caption{Timeline of key invasive speech and motor BCI milestones from 2012 to 2026. The progression highlights a rapid acceleration in speed and channel count, leading to high-dimensional decoding (Willett et al., 2021) and real-time speech avatar synthesis (Metzger et al., 2023).}
\label{fig:invasive_speech_timeline}
\end{figure}

Simultaneously, Willett et al. (2023) \cite{willett2023high} utilizing intracortical arrays in the premotor cortex demonstrated speech decoding at 62 words per minute with a vocabulary of over 100,000 words. Furthermore, Metzger et al. (2024) \cite{metzger2024bilingual} demonstrated a bilingual speech neuroprosthesis, proving that cortical articulatory representations are shared across languages. The trajectory has continued at pace: Card et al. (2024) \cite{card2024accurate} reported a rapidly calibrating intracortical speech neuroprosthesis in a participant with ALS that sustained high accuracy over a large vocabulary, and Littlejohn et al. (2025) \cite{littlejohn2025streaming} demonstrated a streaming brain-to-voice system that synthesized speech near-synchronously with the user's attempt to speak. This efferent pathway represents the most critical avenue for restoring natural communication in patients with anarthria, with systems capable of synthesizing a user's own pre-recorded vocal profile to output fluent, conversational speech at near-natural rates. The rapid progress in clinical validation and performance of these invasive interfaces over the past decade is detailed in the timeline in Figure~\ref{fig:invasive_speech_timeline}.

\subsubsection{Emerging Platforms: Neuralink, Wireless Microelectrodes, and Endovascular Stents}
To overcome the physical limitations of traditional implants, emerging neural interface platforms focus on increasing electrode count, eliminating transcutaneous wires, and reducing surgical trauma. A foundational proof of concept was provided by Vansteensel et al. (2016) \cite{vansteensel2016fully}, who implanted a fully internalized subdural electrode system over the motor cortex of a locked-in patient with ALS, enabling reliable, self-paced communication from home. The Neuralink N1 platform, building on the architecture described by Musk et al. (2019) \cite{musk2019integrated}, represents a major advance in surgical scalability, employing a robotic inserter to distribute more than one thousand recording electrodes across dozens of thin, flexible polymer threads inserted directly into the cortex. The N1 implant is fully wireless and hermetically sealed, eliminating the infection risk of transcutaneous pedestals.

In parallel, minimally invasive approaches such as the Synchron Stentrode utilize endovascular access. The stent-electrode array is guided through the jugular vein into the superior sagittal sinus adjacent to the motor cortex, enabling chronic local field potential recording without open craniotomy. The endovascular approach was established preclinically by Oxley et al. (2016) \cite{oxley2016minimally}, who recorded cortical activity for months from within the vasculature; subsequently, the first-in-human SWITCH study (Mitchell et al., 2023 \cite{mitchell2023assessment}) reported that the device was implanted safely in four patients with severe paralysis, with no serious adverse events related to the device. However, endovascular sensors must record through the blood vessel wall, resulting in lower spatial resolution compared to penetrating microelectrodes. These emerging platforms present a general trade-off: wireless, high-density systems such as Neuralink offer high data throughput with cosmetic invisibility, while endovascular stents minimize surgical risk but limit control bandwidth.

\subsubsection{Biophysical and Clinical Trade-offs: Signal Fidelity vs. Glial Scarring and Surgical Risk}
The clinical translation of invasive BCIs is governed by a strict trade-off between the fidelity of the recorded or stimulated signal and the biophysical response of the brain tissue. Intracortical penetrating electrodes provide unparalleled spatial-temporal resolution by recording action potentials of individual neurons. However, the mechanical mismatch between rigid silicon electrodes and soft brain tissue triggers a chronic inflammatory response. Reactive astrocytes and microglia form a glial sheath (encapsulation) around the electrode contacts, increasing electrical impedance and physically displacing the recording sites from active neurons. This glial response is a primary driver of long-term signal decay, often rendering microelectrode arrays non-functional within 2 to 5 years.

Furthermore, any surgical intervention involving craniotomy carries inherent risks of hemorrhage, infection, and tissue necrosis. Consequently, patients must balance the immediate high-fidelity communication gains against the potential need for revision surgeries. Regulatory pathways, such as the FDA's Investigational Device Exemption (IDE), impose rigorous safety standards, limiting widespread clinical adoption. For an individual patient, the decision to undergo surgical implantation is therefore a complex evaluation of device lifespan, the immediate utility of communication restoration, and the physical risks associated with neurosurgery (Box~\ref{box:anna}).

\subsection{Non-Invasive Modalities}
\label{subsec:noninvasive}
\textit{This subsection charts extracranial acquisition and stimulation modalities, highlighting the role of machine learning in compensating for tissue attenuation.}

\subsubsection{Sensory-OUT: Silent-Speech and Imagined-Speech Decoding}
Non-invasive speech decoding represents one of the most promising frontiers in communication restoration, offering a potential path for paralyzed patients to communicate without undergoing brain surgery. This modality is generally categorized into two paradigms: imagined speech (or covert speech) and silent speech. Imagined speech refers to the internal generation of words without any overt motor activation, which is captured via scalp electroencephalography (EEG) or magnetoencephalography (MEG). In contrast, silent speech involves the generation of speech movements without phonation (subvocal articulation), which can be recorded using surface electromyography (sEMG) sensors placed on the neck and jaw.

\begin{figure}[htbp]
\centering
\includegraphics[width=0.7\textwidth]{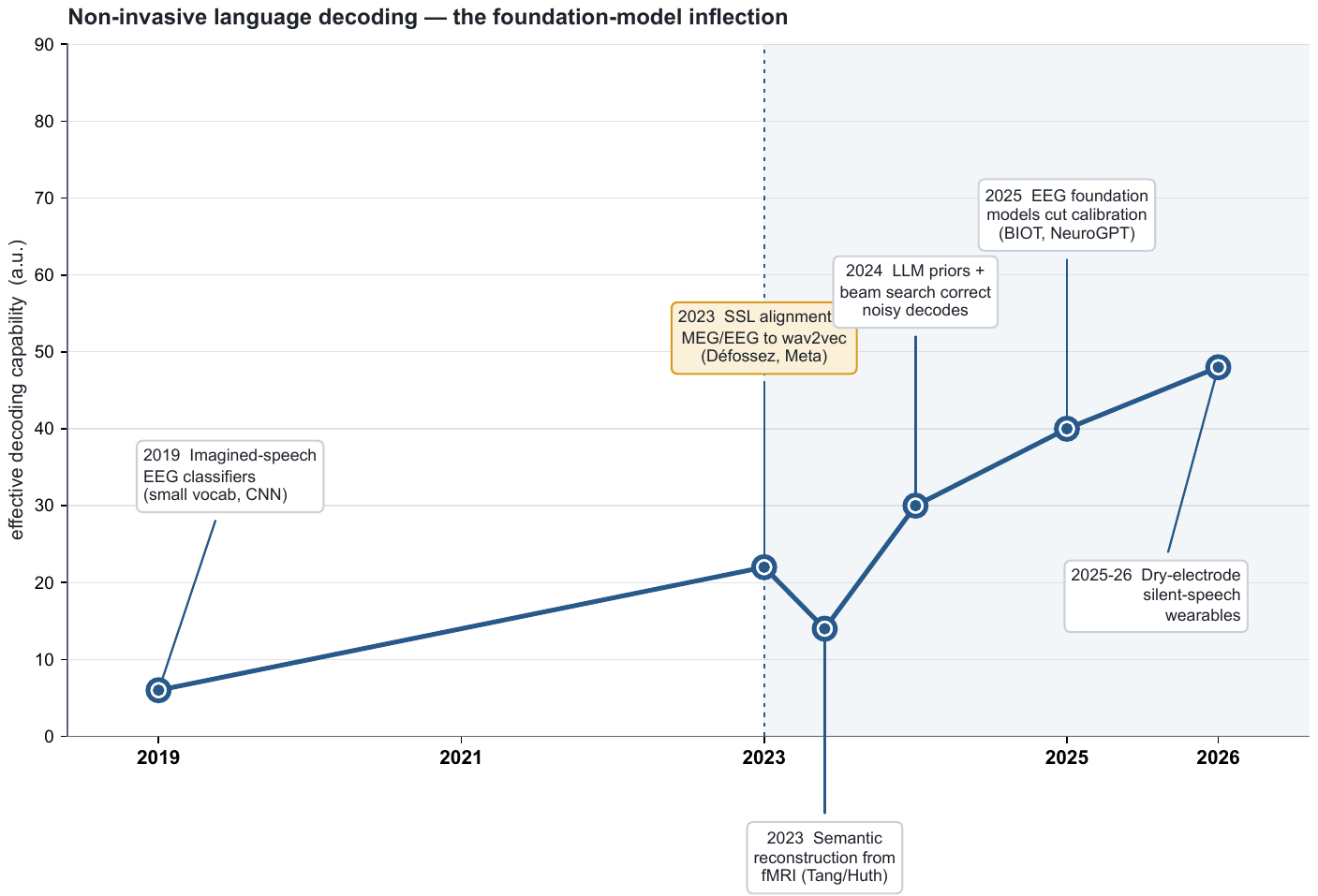}
\caption{Timeline of non-invasive language-decoding milestones from 2019 to 2026. A notable inflection point occurs in late 2023 with the alignment of electrophysiological signals to self-supervised language representations (D\'{e}fossez et al., 2023), enabling robust decoding using large language models.}
\label{fig:noninvasive_speech_timeline}
\end{figure}

The central challenge in non-invasive decoding is the extremely low signal-to-noise ratio (SNR) caused by skull attenuation. To address this, modern systems utilize self-supervised representation learning. A landmark study by D\'{e}fossez et al. (2023) \cite{defossez2023decoding} developed a contrastive learning pipeline that successfully aligned MEG and EEG brain recordings during speech perception with self-supervised speech representations (such as wav2vec 2.0). In the same period, Tang et al. (2023) \cite{tang2023semantic} demonstrated that the semantic content of perceived and imagined speech could be reconstructed as continuous language from non-invasive functional MRI recordings using a generative language model as a prior, underscoring that the principal bottleneck for non-invasive decoding has shifted from signal acquisition to representational alignment. Panachakel \& Ramakrishnan (2021) \cite{panachakel2021decoding} reviewed deep learning methods for covert speech, highlighting the efficacy of spatial-temporal convolutional networks in classifying silent vocabulary. Non-invasive silent-speech BCIs offer an immediate, low-risk communication channel: setup avoids the surgical risks of craniotomy, though decoding speed and accuracy may be lower due to signal attenuation. The evolution and recent machine learning inflection points of non-invasive language decoding are summarized in the timeline in Figure~\ref{fig:noninvasive_speech_timeline}.

\subsubsection{Sensory-OUT: Motor Imagery and Assistive Device Control}
Motor imagery (MI) paradigms rely on the user's ability to modulate sensorimotor rhythms (SMRs) by imagining the movement of specific body parts (e.g., left hand, right hand, or feet). These imaginations elicit event-related desynchronization (ERD) or event-related synchronization (ERS) in the alpha/mu (8--12~Hz) and beta (13--30~Hz) frequency bands over the primary motor cortex.

These oscillations are recorded via scalp EEG and processed using spatial filtering algorithms, such as Common Spatial Patterns (CSP) or Filter Bank Common Spatial Patterns (FBCSP), to isolate the lateralized neural activity. The decoded states are mapped to control interfaces, such as wheelchair navigation commands or visual keyboard cursors. For patients who have lost all voluntary movement, motor imagery can still provide a reliable, non-invasive command channel for simple binary or multi-class control switches.

\subsubsection{Sensory-OUT: Evoked Potential Paradigms (P300, SSVEP)}
Evoked potential paradigms rely on visual attention to external stimuli rather than internally generated motor imagery. The two most common paradigms are the P300 wave and Steady-State Visually Evoked Potentials (SSVEP). The P300 paradigm utilizing the Farwell-Donchin matrix flashes rows and columns of a character grid. When the target letter flashes, it triggers an event-related potential (ERP) characterized by a positive voltage deflection approximately 300~ms post-stimulus.

SSVEP-based BCIs utilize visual targets flickering at distinct frequencies (e.g., 8 to 15~Hz). When the user gazes at a target, the visual cortex exhibits oscillations at the corresponding stimulus frequency, which are decoded using Canonical Correlation Analysis (CCA). While both systems are highly reliable and require minimal user training, they impose a high cognitive and visual workload. The constant flickering can cause severe visual fatigue and headaches over extended use. P300 and SSVEP spellers represent robust backup communication tools, providing a stable spelling option when higher-dimensional speech-decoding models fail or require re-calibration.

\subsubsection{Sensory-IN: Cross-Modal Sensory Substitution Systems}
When primary sensory pathways are destroyed, non-invasive sensory substitution systems route the environmental data through alternative sensory channels. The principle dates to the pioneering tactile-vision system of Bach-y-Rita et al. (1969) \cite{bachyrita1969vision}, who rendered a camera image as a vibrotactile pattern on the skin and showed that blind users could learn to recognize objects through it. In sensory-IN substitution, visual or auditory information is mapped to tactile stimulation on the skin or auditory sonification. Representative contemporary systems include the BrainPort, which translates camera images into electrical stimulation patterns on the tongue, and haptic vests (such as Neosensory) that convert audio frequencies into vibration arrays across the torso.

The biological mechanism driving substitution is cross-modal neuroplasticity, where the brain repurposes the visual or auditory cortex to interpret tactile or acoustic patterns. However, these systems are constrained by the physical limits of the proxy receptor organs. The spatial density of mechanoreceptors in the skin is orders of magnitude lower than the photoreceptor density in the retina, limiting the resolution of tactile-based sight. Such systems can translate ambient sound or visual layouts into tactile vibration patterns, allowing users to perceive the spatial presence of people or objects in their environment.

\subsubsection{Sensory-IN: Non-Invasive Neuromodulation (tFUS, tACS, tDCS)}
Non-invasive neuromodulation uses external magnetic, electric, or acoustic fields to alter cortical excitability without surgical implants. Transcranial Focused Ultrasound (tFUS) uses low-intensity focused ultrasound to mechanically modulate ion channels in targeted brain regions with sub-millimeter precision; Legon et al. (2014) \cite{legon2014transcranial} first showed in humans that tFUS directed at the primary somatosensory cortex could alter sensory-evoked cortical responses and tactile discrimination. Transcranial alternating current stimulation (tACS) and direct current stimulation (tDCS) apply low-level electrical currents via scalp electrodes to modulate membrane potentials.

The primary limitation of non-invasive neuromodulation is spatial resolution. While tFUS achieves millimeter-scale targeting, electrical stimulation methods (tACS/tDCS) are highly diffuse, stimulating large regions of the cortex. Consequently, they cannot elicit localized sensory perceptions (such as individual phosphenes). However, they are highly effective at modulating general cortical excitability, enhancing cognitive performance, or treating depressive states. In patients with prolonged immobility, non-invasive neuromodulation targeting motor regions has been proposed as a means of maintaining cortical excitability and mitigating disuse-related neural decay.

\subsubsection{The Role of Large Language Models and Foundation Models in Electrophysiological Decoding}
The integration of deep learning and language foundation models has revolutionized non-invasive BCI performance. The primary obstacle of EEG and MEG---low spatial-temporal resolution---is bypassed by using neural network foundation models pre-trained on massive datasets across hundreds of subjects. These models (such as BIOT or NeuroGPT) learn subject-agnostic features of electrophysiological signals, reducing the calibration time required for a new patient from hours to minutes.

Furthermore, large language models (LLMs) act as powerful linguistic priors during the decoding phase. Raw, noisy predictions from an EEG silent-speech decoder are passed through an LLM that performs beam search decoding, correcting spelling and grammatical errors based on contextual probabilities. This integration is crucial for clinical usability: rather than relying on slow, error-prone raw-signal decoding, such systems predict entire words or sentences, accelerating communication rates while reducing the cognitive fatigue associated with letter-by-letter spelling.

\subsubsection{Engineering Trade-offs: Scalability and Usability vs. Skull Attenuation and Artifacts}
Non-invasive BCIs represent a trade-off between practical accessibility and signal fidelity. The primary advantages of non-invasive systems are their low cost, rapid deployment, and complete safety. However, they are fundamentally limited by skull attenuation, which filters out high-frequency gamma oscillations and scatters spatial details. Non-invasive recordings are also highly susceptible to muscle artifacts, ocular movements, and eye blinks, requiring sophisticated artifact-rejection pipelines.

Furthermore, the physical preparation of wet EEG caps (requiring conductive gel) is time-consuming and uncomfortable for daily use, while dry electrodes suffer from higher impedance and signal noise. Overall, non-invasive interfaces offer a highly scalable and safe entry point into communication restoration, though daily setup time and limited control bandwidth remain significant adoption barriers.

\subsection{Cross-Pathway Comparison}
\label{sec:comparison}

\begin{figure}[htbp]
\centering
\includegraphics[width=0.8\textwidth]{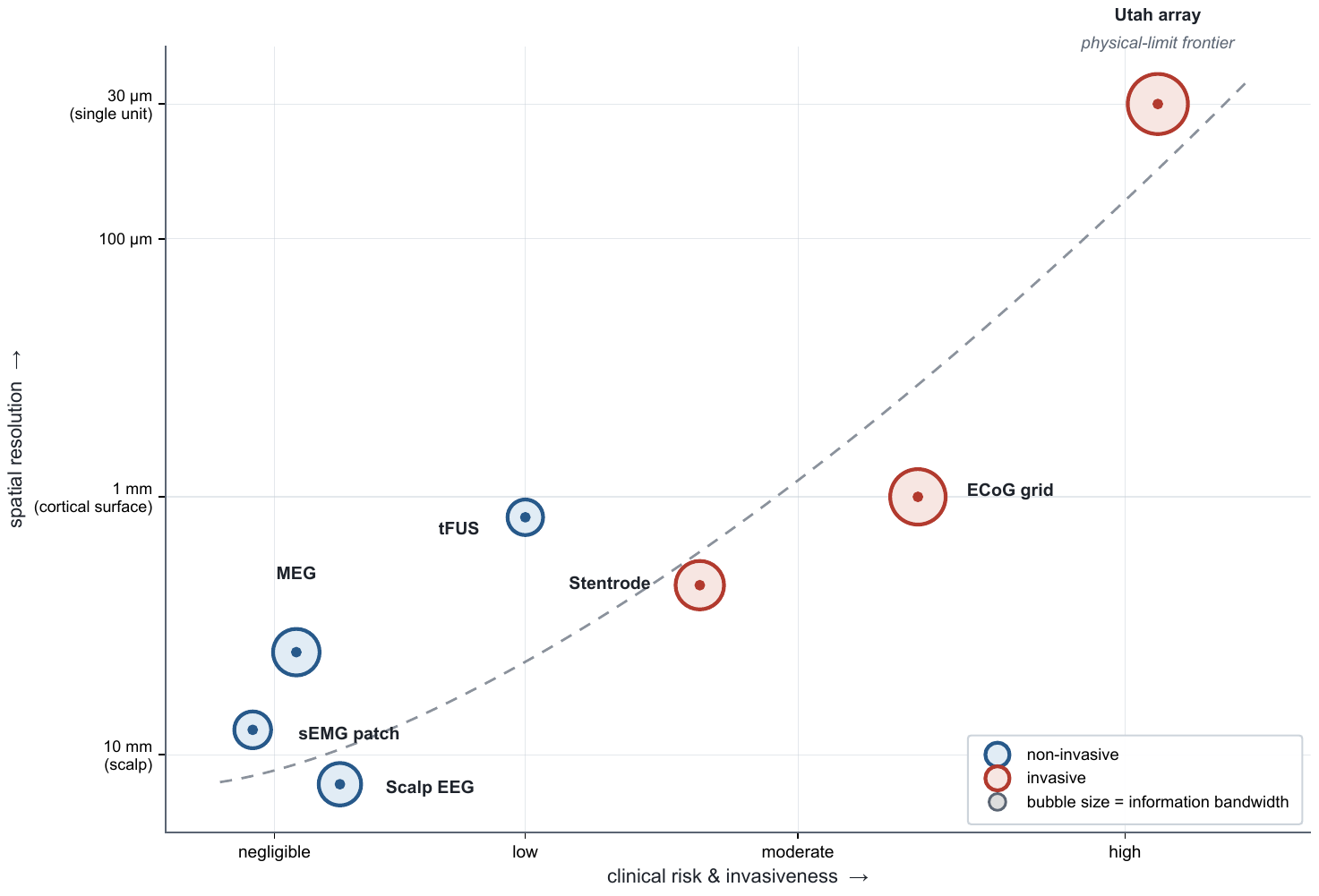}
\caption{Visual comparison of cross-pathway trade-offs. The scatter plot maps different BCI modalities along two primary axes: spatial-temporal resolution (representing signal capacity) versus clinical risk and relative system cost. This highlights the trade-offs described in Table~\ref{tab:trade_offs} and visualizes the physical limits frontier.}
\label{fig:comparison_plot}
\end{figure}

To evaluate the parallel routes of sensory restoration, we compare the representative modalities across clinical, biological, and engineering dimensions. Table~\ref{tab:trade_offs} summarizes these trade-offs side by side. To visually map these parameters, Figure~\ref{fig:comparison_plot} plots the spatial resolution of each interface against its associated clinical risk and relative cost, illustrating the physical limits frontier that researchers must navigate.

\begin{table}[htbp]
\centering
\caption{Cross-Pathway Technical and Clinical Trade-offs}
\label{tab:trade_offs}
\vspace{1mm}
\resizebox{\linewidth}{!}{%
\begin{tabular}{llllll}
\toprule
\textbf{Modality} & \textbf{Signal Type} & \textbf{Spatial Res.} & \textbf{Clinical Risk} & \textbf{Relative Cost} & \textbf{Maturity} \\
\midrule
Utah Array & Single-unit Spikes & High ($\approx 10\,\mu\text{m}$) & High (Glial Scarring) & High & Investigational \\
ECoG Grid & Local Field Potentials & Med ($\approx 1\,\text{mm}$) & Moderate & High & Investigational \\
Stentrode & Venous LFP & Med ($\approx 2\,\text{mm}$) & Low-Moderate & Moderate & Early Clinical \\
Scalp EEG & Scalp Volts & Low ($\approx 10\,\text{mm}$) & Negligible & Low & Commercial \\
MEG & Magnetic Fields & Low-Med ($\approx 5\,\text{mm}$) & Negligible & Very High & Research \\
sEMG Patch & Muscle Volts & Med ($\approx 5\,\text{mm}$) & Negligible & Very Low & Commercial \\
tFUS & Acoustic Waves & Med ($\approx 1\,\text{mm}$) & Low (Thermal Risk) & Moderate & Research \\
\bottomrule
\end{tabular}%
}
\end{table}

Clinical decision-making for sensory restoration must weigh surgical tolerance, disease progression rate, and communication requirements. For a patient in the early stages of a neurodegenerative disease who retains residual motor function and desires zero surgical risk, a non-invasive scalp EEG or silent-speech sEMG system is typically indicated. For a patient in late-stage ALS or a locked-in state (Box~\ref{box:anna}), where cognition is intact but motor output is fully blocked, the desire for high-speed, natural communication may justify the risks of a surgical implant; a high-density sub-dural ECoG grid often offers the optimal balance between signal bandwidth and tissue damage, consistent with the trade-offs in Table~\ref{tab:trade_offs}.

\section{Discussion}
\label{sec:discussion}

\subsection{Towards Convergence: A Phased Roadmap}
\label{subsec:roadmap}
\begin{figure}[htbp]
\centering
\includegraphics[width=0.9\textwidth]{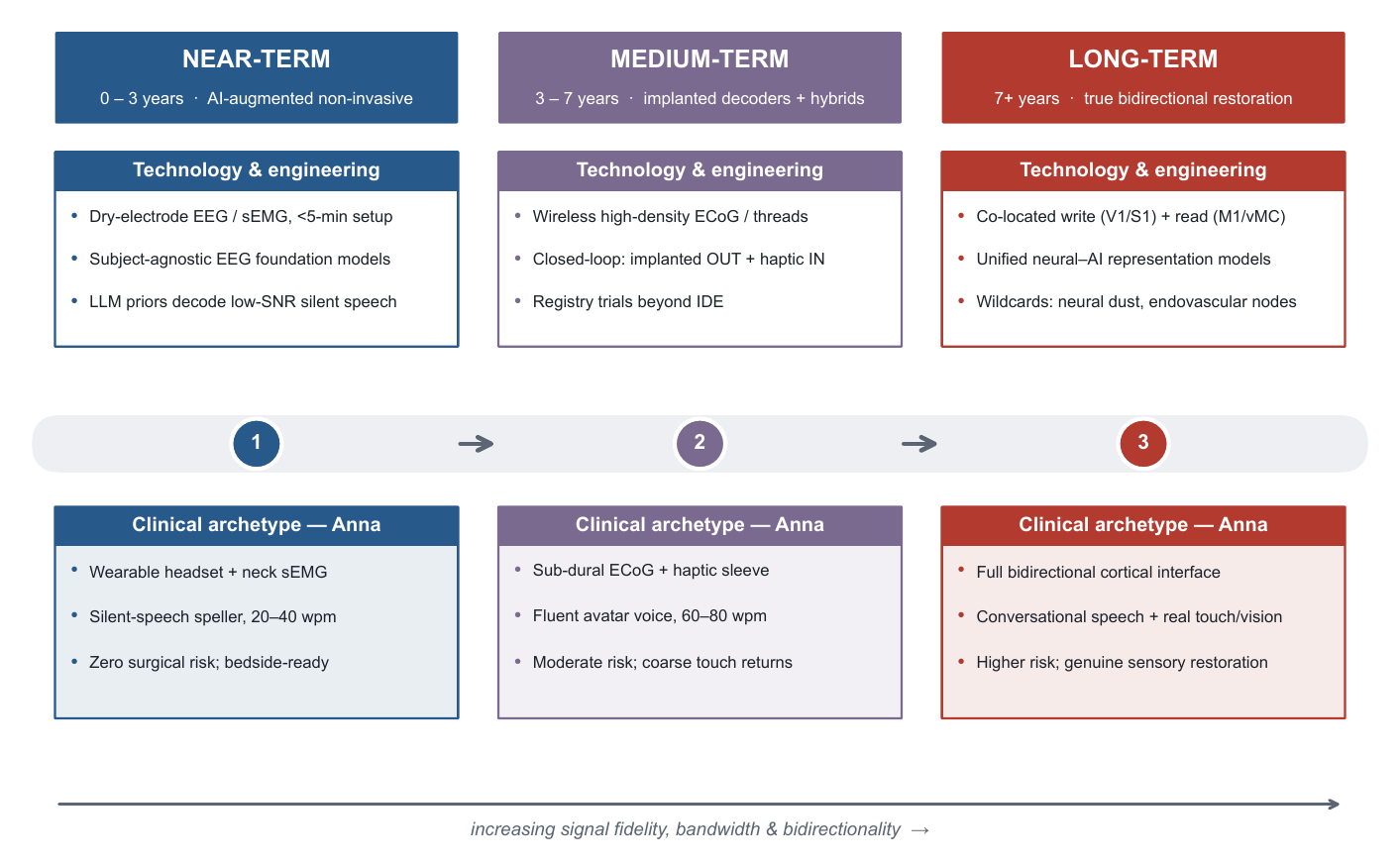}
\caption{BCI convergence roadmap. Development proceeds through three phases: (1) near-term non-invasive communication, (2) medium-term implanted high-density decoders and closed-loop hybrid systems, and (3) long-term true bidirectional restoration, combining S1/V1 microstimulation and speech motor decoding.}
\label{fig:convergence_roadmap}
\end{figure}

\textit{This subsection projects the evolution and ultimate integration of invasive and non-invasive modalities. The multi-phase convergence roadmap spanning near-term non-invasive systems to long-term bidirectional, AI-fused cortical interfaces is illustrated in Figure~\ref{fig:convergence_roadmap}.}

\subsubsection{Near-Term (0--3 Years): Non-Invasive, AI-Augmented Communication}
In the immediate future, communication and sensory restoration will be dominated by non-invasive electrophysiological and biomechanical systems. The rapid advancement of large language models (LLMs) and subject-independent neural network pre-training will make non-invasive silent-speech and imagined-speech systems highly practical. These devices will achieve typing speeds of 20 to 40 words per minute using scalp EEG or neck sEMG sensors, with setup times dropping below 5 minutes due to dry-electrode innovations. For patients in a locked-in state (Box~\ref{box:anna}), this near-term horizon provides an accessible, non-surgical communication link that could be deployed at the bedside immediately, restoring the ability to express basic thoughts, emotions, and physical needs without surgical intervention.

\subsubsection{Medium-Term (3--7 Years): Implanted High-Density Decoders and Closed-Loop Hybrids}
The medium-term horizon will witness the clinical expansion of wireless, high-density neural interfaces. Implanted platforms like the Neuralink N1 or the Synchron Stentrode will move from investigational device exemptions (IDE) to broader clinical registry trials, offering permanent, high-bandwidth communication links. We will also see the emergence of closed-loop hybrid systems that combine pathways: using implanted ECoG grids for efferent speech decoding (sensory-OUT) and non-invasive haptic vests or tactile arrays for sensory feedback (sensory-IN). For patients whose non-invasive signals degrade with disease progression (e.g., muscle atrophy in ALS; Box~\ref{box:anna}), this period represents a potential transition to a wireless, sub-dural ECoG grid providing high-fidelity speech reconstruction directly mapped to a digital avatar.

\subsubsection{Long-Term (7+ Years): True Bidirectional Visual and Somatosensory Restoration}
\begin{figure}[htbp]
\centering
\includegraphics[width=0.95\textwidth]{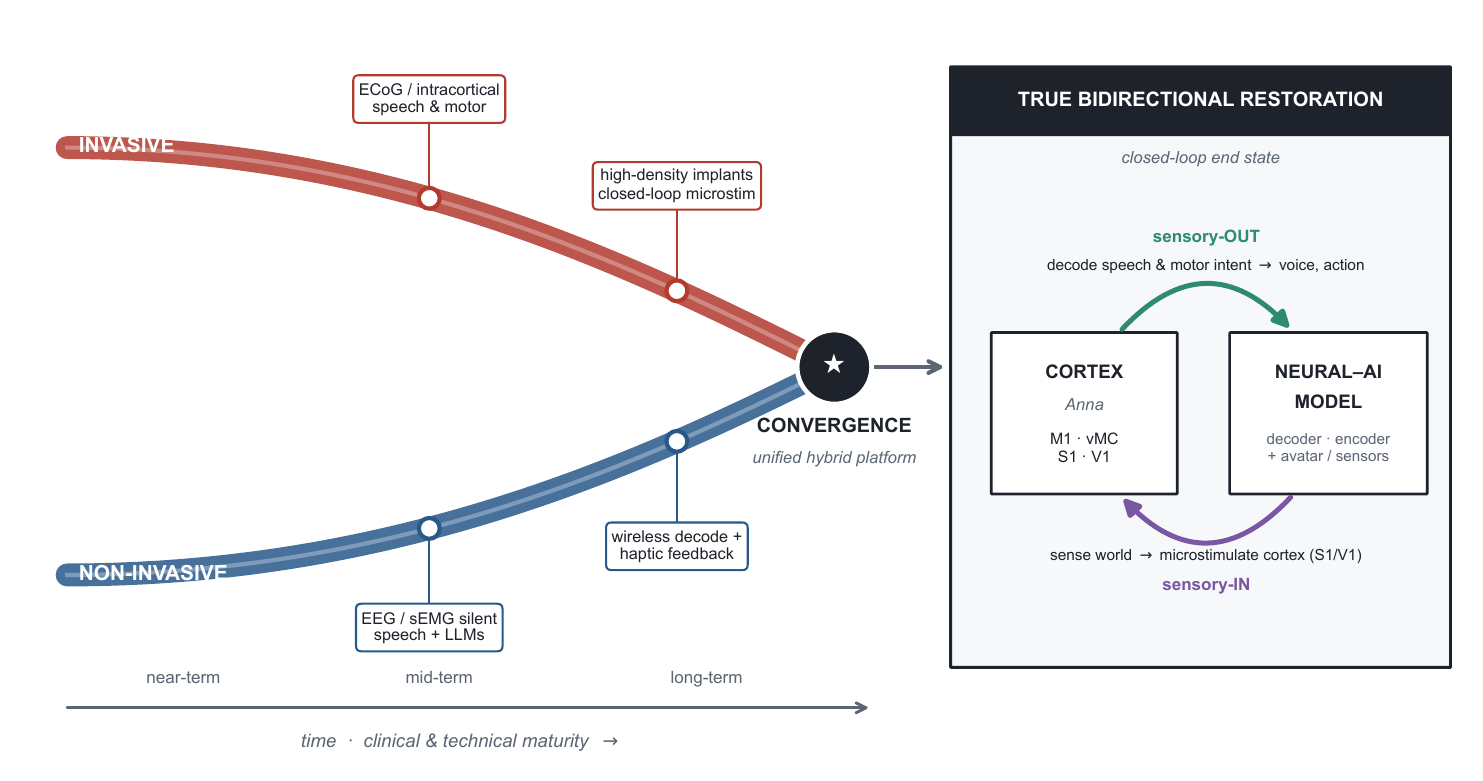}
\caption{Convergence of the two pathways into a closed-loop end state. As clinical and technical maturity increase (left to right), the invasive (red) and non-invasive (blue) pathways progress through their respective near-, mid-, and long-term milestones and converge into a unified hybrid platform. The end state (right) is a closed-loop bidirectional system in which a shared neural--AI model couples efferent decoding of speech and motor intent (sensory-OUT) with afferent cortical microstimulation of S1/V1 (sensory-IN), reconnecting the user to both communication and sensation.}
\label{fig:convergence_endstate}
\end{figure}

Over the long term, BCI development will converge on true bidirectional restoration. This phase will be characterized by fully closed-loop systems capable of recording high-dimensional motor intent while simultaneously writing sensory information back to the cortex. This will require microstimulation arrays (in V1 and S1) to work in synchronization with recording arrays (in M1 and vMC) under the control of unified machine learning models that translate stimulation profiles to align with the brain's internal representations. This long-term integration represents the ultimate goal of the field: systems that not only restore fluent communication through a digital avatar but also genuine, high-resolution touch and visual feedback, reconnecting users to their physical and social environment (Box~\ref{box:anna}). Figure~\ref{fig:convergence_endstate} depicts how the two pathways, having matured through the phases described above, ultimately merge into a single hybrid platform whose end state is a closed-loop bidirectional architecture---one in which efferent decoding (sensory-OUT) and afferent microstimulation (sensory-IN) operate as a single continuous loop mediated by a shared neural--AI model.

\subsubsection{Technological Wildcards: Nanoscale Neural Interfaces and Endovascular Nodes}
Beyond standard silicon and polymer arrays, several technological wildcards may disrupt the BCI field. These include nanoscale neural dust---microscopic wireless sensors scattered throughout the cortex that communicate via ultrasound---and injectable endovascular nodes that cross the blood-brain barrier to record or stimulate localized cortical columns without open craniotomy. While still in early pre-clinical development, these nanoscale interfaces could bypass the glial response and surgical risks that currently limit intracortical microelectrodes. Should these wildcards transition to clinical viability, they could offer a less invasive route to high-bandwidth neural restoration, eliminating long-term concerns regarding device degradation or tissue scarring.

\subsection{Critical Gaps in the Contemporary Literature}
\label{subsec:gaps}
Charting the literature across both pathways reveals three major gaps that restrict clinical translation:
\begin{enumerate}
    \item \textbf{Metric Discontinuity:} The invasive and non-invasive research communities employ divergent evaluation metrics. Efferent speech studies report Word Error Rates (WER) and Character Error Rates (CER), whereas afferent stimulation studies report phosphene counts or sensory thresholds. This lack of standardization prevents direct comparison of communication efficiency.
    \item \textbf{Longitudinal Stability:} There is a lack of long-term longitudinal studies evaluating the chronic performance of both pathways. For invasive systems, studies rarely report signal degradation beyond 2 years; for non-invasive systems, the daily variability in electrode placement and gel dehydration remains unstudied.
    \item \textbf{Community Siloing:} Research remains siloed. Teams developing high-density invasive stimulation interfaces rarely collaborate with teams building non-invasive sensory substitution setups, leaving a gap in hybrid closed-loop designs.
\end{enumerate}

\subsection{Ethical, Agency, and Regulatory Considerations}
\label{subsec:ethics}
\textit{This subsection addresses the philosophical and socio-regulatory implications of modifying human perception.}

\subsubsection{Neuro-Agency and Autonomy: Who Controls a Synthesized Sense?}
Restoring or substituting human senses using machine learning decoders introduces profound questions regarding user agency. When an algorithm processes visual or auditory inputs---filtering, enhancing, or selecting which spatial features to stimulate---the user's perception is mediated by a computational middle layer. If a visual prosthesis uses computer vision to highlight objects or filter out details to prevent phosphene overlap, the system is actively deciding what the user perceives. Defining the boundaries of agency is crucial: we must ensure that the BCI acts as a transparent tool of the user's will rather than an autonomous actor that shapes their subjective experience. These concerns have been crystallized in proposals for a dedicated set of ``neurorights''---safeguarding privacy, identity, agency, and equality---articulated by Yuste et al. (2017) \cite{yuste2017four} as ethical priorities that must precede the widespread deployment of neurotechnology.

If a speech BCI auto-completes an utterance that diverges from the user's actual intent (e.g., predicting ``I need water'' when the user thought ``I need rest''; Box~\ref{box:anna}), the user's agency is compromised. The user must remain in control of the final output channel, and language-model predictive weights must be adjustable to ensure genuine autonomous expression.

\subsubsection{Cognitive Privacy: Protecting Raw EEG/ECoG Speech Intention Data}
Electrophysiological data is the most private form of biological information, capturing the raw, pre-conscious intentions of the user. Speech neuroprosthetics record covert linguistic representations, meaning they access unspoken thoughts and inner monologues. If this raw data is transmitted to external cloud servers for machine learning decoding, it creates a severe risk of cognitive privacy violations. Unintended disclosures of private thoughts, emotions, or memories could occur. Protecting this data requires the development of secure, on-device edge computing models and strict regulatory frameworks that classify neural signals as highly protected medical data.

In practice, raw brain signals should be processed locally on bedside or wearable hardware to prevent data leaks, and transmission of raw neural data over public networks to third-party language models should be prohibited.

\subsubsection{Informed Consent in Locked-In Patient Populations}
Establishing authentic informed consent is a major challenge in clinical trials involving locked-in patients. Because these patients have lost standard communication channels, ensuring they comprehend the high surgical risks, potential failure modes, and long-term maintenance requirements of invasive implants is difficult. Researchers must utilize reliable non-invasive communication backups (such as P300 spellers) to conduct rigorous, multi-stage consent protocols, ensuring that the patient's decision is autonomous and free from external coercion.

In practice, a patient's existing non-invasive communication channel (e.g., a P300 speller; Box~\ref{box:anna}) can be used by caregivers and clinical teams to run structured consent questionnaires ahead of implant surgery, verifying that the patient fully understands the risk-to-benefit profile.

\subsubsection{Regulatory Pathways and Safety Standards (FDA, CE Mark)}
The translation of BCIs from research labs to clinical markets is governed by regulatory safety standards, such as the FDA's Premarket Approval (PMA) and the European Union's Medical Device Regulation (MDR). These frameworks impose strict guidelines on biocompatibility, wireless telemetry heating, and electromagnetic compatibility. Meeting these standards is financially and logistically demanding, creating a commercialization bottleneck that often prevents promising technologies from reaching patients.

In practice, a patient's choice of implant is restricted to devices currently holding FDA Investigational Device Exemption (IDE) approval or equivalent clearance for clinical trials, often requiring patients to wait years for advanced models.

\section{Limitations}
\label{sec:limitations}
As a scoping review, this work prioritizes breadth of mapping over exhaustive coverage or quality/risk-of-bias appraisal of individual studies, consistent with PRISMA-ScR guidance. Several limitations should be considered when interpreting the synthesis in \S\ref{sec:results}--\S\ref{sec:discussion}:
\begin{itemize}
    \item \textbf{Purposive, non-exhaustive search:} This review does not perform a comprehensive multi-database search. Sources were identified by citation chaining from seminal milestones and deliberately restricted to high-impact venues (\S\ref{subsec:search}). This anchors the synthesis in well-validated, landmark evidence but introduces a clear selection bias: incremental and negative results, work from less-resourced groups, preprints, non-indexed conference proceedings, and the substantial non-invasive and afferent literature published in specialist venues are systematically under-represented (cf. the 22/31 invasive and 21/31 efferent skew in Table~\ref{tab:distribution}). The reported distributions therefore characterize the \textit{landmark} evidence base, not the field as a whole, and a future update employing a full systematic search would refine them.
    \item \textbf{Language restriction:} Only English-language publications were considered, which may bias the evidence base against relevant work published in other languages and against research output from non-anglophone regions.
    \item \textbf{Rapidly evolving field:} Benchmark performance in both invasive and non-invasive speech/motor decoding has advanced substantially year-over-year (Figures~\ref{fig:invasive_speech_timeline}--\ref{fig:noninvasive_speech_timeline}); numeric milestones cited here may be superseded before publication. We mitigate this by emphasizing durable structural principles (\S\ref{subsec:taxonomy}) over point-in-time benchmarks.
    \item \textbf{Heterogeneous reporting:} As noted in \S\ref{subsec:gaps}, the absence of standardized metrics across afferent and efferent studies limits the precision of the cross-pathway comparison in Table~\ref{tab:trade_offs} and Figure~\ref{fig:comparison_plot}, which should be read as indicative rather than statistically pooled estimates.
    \item \textbf{Single-reviewer charting and no registered protocol:} Screening and data charting (\S\ref{subsec:selection}--\S\ref{subsec:charting}) were performed by a single reviewer without independent verification or a pre-registered protocol, increasing the risk of selection and extraction bias. The qualitative quadrant and modality assignments in Table~\ref{tab:charted} reflect one reviewer's interpretation and would benefit from dual-reviewer adjudication in a future iteration.
\end{itemize}

\section{Conclusion}
\label{sec:conclusion}

\subsection{Summary of the $2\times2$ Taxonomy}
The $2\times2$ framework applied in this review provides a unified taxonomy that bridges the historically fragmented fields of invasive and non-invasive BCIs. By mapping modalities along the axes of invasiveness and information flow, researchers and clinicians can systematically evaluate the spatial-temporal bandwidth, safety risks, and regulatory barriers of any BCI system. This structured perspective helps align terminology across the diverse scientific and engineering sub-disciplines involved in neuroprosthetic design. For an individual patient (Box~\ref{box:anna}), this taxonomy maps clinical options onto a single logical grid, helping patients, families, and clinicians visualize how different devices fit a given disease stage and personal risk tolerance.

\subsection{Illustrative Synthesis (Box~\ref{box:anna})}
Returning to the illustrative vignette in Box~\ref{box:anna}, the $2\times2$ framework maps a clear, progressive clinical pathway: in the immediate term, a non-invasive, AI-augmented silent-speech BCI can restore communication without surgical risk; as disease progresses, transition to a wireless, high-density ECoG implant can achieve fluent, conversational speech; and in the long term, convergence toward bidirectional cortical stimulation offers a pathway to restore touch and spatial presence---illustrating how these parallel technical routes merge to restore human connection.

\subsection{A Call for Interdisciplinary Collaboration}
Realizing the full potential of sensory neuroprosthetics demands breaking down the siloes between material science, neurosurgery, machine learning, and bioethics. Only through interdisciplinary collaboration can we design BCI systems that are biocompatible, computationally efficient, and ethically sound, ultimately returning agency and sensory richness to the millions of individuals who have been locked away from the world.

A commitment to ethical design, unified standards, and cross-pathway communication---together with future updates to this scoping review as the evidence base matures---is necessary to translate these engineering milestones into clinical realities, ensuring that patients with severe sensory and motor impairment are never again left without a voice or a sense.

\bibliographystyle{plain}
\bibliography{references}

\end{document}

%% file: charted_sources_table.tex
\begin{footnotesize}
\begin{longtable}{@{}p{0.20\textwidth}c p{0.15\textwidth} p{0.26\textwidth} c p{0.10\textwidth}@{}}
\caption{Characteristics of the 31 charted sources, ordered by year. Direction: afferent (sensory-IN), efferent (sensory-OUT), or cross-cutting (reviews/ethics).}\label{tab:charted}\\
\toprule
\textbf{Source} & \textbf{Yr} & \textbf{Venue} & \textbf{Modality / platform} & \textbf{Dir.} & \textbf{Type} \\
\midrule
\endfirsthead
\multicolumn{6}{c}{\textit{Table~\ref{tab:charted} (continued)}}\\
\toprule
\textbf{Source} & \textbf{Yr} & \textbf{Venue} & \textbf{Modality / platform} & \textbf{Dir.} & \textbf{Type} \\
\midrule
\endhead
\bottomrule
\endfoot
\cite{bachyrita1969vision} & 1969 & Nature & Tactile--vision substitution & Aff. & Primary \\
\cite{wolpaw2002brain} & 2002 & Clin. Neurophysiol. & BCI communication \& control & Cross & Review \\
\cite{hochberg2006neuronal} & 2006 & Nature & Utah array (M1 cursor) & Eff. & Primary \\
\cite{hochberg2012reach} & 2012 & Nature & Utah array, robotic arm & Eff. & Primary \\
\cite{collinger2013high} & 2013 & Lancet & Utah array, 10-DoF arm & Eff. & Primary \\
\cite{legon2014transcranial} & 2014 & Nat. Neurosci. & tFUS neuromodulation (S1) & Aff. & Primary \\
\cite{gilja2015clinical} & 2015 & Nat. Med. & Intracortical decoder & Eff. & Primary \\
\cite{aflalo2015decoding} & 2015 & Science & Posterior parietal array & Eff. & Primary \\
\cite{flesher2016intracortical} & 2016 & Sci. Transl. Med. & S1 ICMS (touch) & Aff. & Primary \\
\cite{vansteensel2016fully} & 2016 & NEJM & Implanted ECoG (LIS/ALS) & Eff. & Primary \\
\cite{bouton2016restoring} & 2016 & Nature & Intracortical + FES bypass & Eff. & Primary \\
\cite{oxley2016minimally} & 2016 & Nat. Biotechnol. & Endovascular Stentrode & Eff. & Primary \\
\cite{chaudhary2016brain} & 2016 & Nat. Rev. Neurol. & BCI in paralysis & Cross & Review \\
\cite{lebedev2017brain} & 2017 & Physiol. Rev. & Brain--machine interfaces & Cross & Review \\
\cite{yuste2017four} & 2017 & Nature & Neuroethics / neurorights & Cross & Perspective \\
\cite{anumanchipalli2019speech} & 2019 & Nature & ECoG speech synthesis & Eff. & Primary \\
\cite{musk2019integrated} & 2019 & JMIR & Neuralink N1 platform & Eff. & Primary \\
\cite{beauchamp2020dynamic} & 2020 & Cell & V1 dynamic stimulation & Aff. & Primary \\
\cite{fernandez2021visual} & 2021 & J. Clin. Invest. & V1 Utah array (phosphenes) & Aff. & Primary \\
\cite{willett2021high} & 2021 & Nature & Intracortical handwriting & Eff. & Primary \\
\cite{moses2021neuroprosthesis} & 2021 & NEJM & ECoG speech (anarthria) & Eff. & Primary \\
\cite{flesher2021brain} & 2021 & Science & S1 ICMS + robotic arm & Aff. & Primary \\
\cite{panachakel2021decoding} & 2021 & Front. Neurosci. & Covert-speech EEG & Eff. & Review \\
\cite{metzger2023high} & 2023 & Nature & ECoG speech avatar & Eff. & Primary \\
\cite{willett2023high} & 2023 & Nature & Intracortical speech & Eff. & Primary \\
\cite{defossez2023decoding} & 2023 & Nat. Mach. Intell. & MEG/EEG speech decoding & Eff. & Primary \\
\cite{tang2023semantic} & 2023 & Nat. Neurosci. & fMRI semantic decoding & Eff. & Primary \\
\cite{mitchell2023assessment} & 2023 & JAMA Neurol. & Stentrode (SWITCH trial) & Eff. & Primary \\
\cite{metzger2024bilingual} & 2024 & Nat. Biomed. Eng. & ECoG bilingual speech & Eff. & Primary \\
\cite{card2024accurate} & 2024 & NEJM & Intracortical speech (ALS) & Eff. & Primary \\
\cite{littlejohn2025streaming} & 2025 & Nat. Neurosci. & ECoG streaming brain-to-voice & Eff. & Primary \\
\end{longtable}
\end{footnotesize}